\documentclass[english,cits]{PoS}
\pdfoutput=1
\usepackage[utf8]{inputenc}
\usepackage[numbers,square]{natbib}
\usepackage{caption}
\usepackage{array}
\usepackage{setspace}

\title{The strange contribution to $a_{\mu}$ with physical quark masses
using Möbius domain wall fermions}

\ShortTitle{Strange contribution to $a_\mu$ using M\"obius domain wall fermions}

\author{\speaker{Matt Spraggs}$^a$, Peter Boyle$^b$, Luigi Del Debbio$^b$, Andreas J\"uttner$^a$, Christoph Lehner$^c$, Kim Maltman$^{d,e}$, Marina Marinkovic$^f$, Antonin Portelli$^{a,c}$\\
  $^a$School of Physics and Astronomy, University of Southampton, Southampton SO17 1BJ, UK\\
  $^b$SUPA, School of Physics, The University of Edinburgh, Edinburgh EH9 3JZ, UK\\
  $^c$Physics Department, Brookhaven National Laboratory, Upton, NY 11973, US\\
  $^d$Department of Physics and Astronomy, York University, Toronto, Ontario, M3J 1P3, Canada\\
  $^e$CSSM, University of Adelaide, Adelaide, SA 5005, Australia\\
  $^f$CERN, Physics Department, 1211 Geneva 23, Switzerland\\
  Email: \email{ms10g12@soton.ac.uk}}
\abstract{
  We present preliminary results for the strange leading-order hadronic contribution
  to the anomalous magnetic moment of the muon using RBC/UKQCD physical point domain
  wall fermions ensembles. We discuss various analysis strategies in order to
  constrain the systematic uncertainty in the final result.
}

\FullConference{The 33rd International Symposium on Lattice Field Theory,\\
		14-18 July, 2015\\
		Kobe International Conference Center, Kobe, Japan}

\usepackage{babel}
\begin{document}

\setstretch{1.02}

\section{Introduction}

The anomalous magnetic moment of the muon, $a_\mu$, is one of the most
accurately determined quantities in particle physics, with an accuracy
of the order of one part per million \cite{Jegerlehner:2009ry}. There
is currently a 3$\sigma$ to 4$\sigma$ tension between the experimental
and theoretical determinations of this quantity.  The new muon $g-2$
experiment at Fermilab is expected to reduce the uncertainty from
experiment by a factor of four, making a reduction in the theoretical
error desirable.  The leading-order (LO) hadronic contribution is the
main source of this uncertainty. In addition, current estimates of
this value are computed using a
$\sigma\left(e^+ e^-\rightarrow {\rm hadrons}\right)$ data
\cite{Davier:2010nc,Hagiwara:2011af}, making a first-principles
computation desirable.  Here we present the computation of the
connected strange contribution to this quantity. We use a variety of
analysis techniques in order to test both the techniques and their
effect on the final value of $a_\mu^{\rm s}$.

The LO strange hadronic contribution, $a_\mu^{\rm s}$, can be computed
as follows \cite{Blum:2002ii}:

\begin{equation}
  \label{eq:amu-integral}
  a_{\mu}^{\rm s}=\left(\frac{\alpha}{\pi}\right)^2\int_{0}^{\infty}{\rm d}Q^{2} \hat{\Pi}\left(Q^2\right)f\left(Q^{2}\right),
\end{equation}

\noindent
where $\alpha$ is the QED coupling,
$\hat{\Pi}(Q^2)=4\pi^2\left(\Pi^{\rm s}(Q^2) - \Pi^{\rm s}(0)\right)$
is the infra-red subtracted hadronic vacuum polarization (HVP) scalar
function and $f$ is the integration kernel derived in perturbation
theory, with a singularity at $Q^{2}=0$. The resulting integrand is
highly peaked near $Q^2\approx m_\mu^2 / 4$, meaning that the final
value of $a_\mu$ is highly sensitive to variations in the values of
$\hat{\Pi}(Q^2)$.

Our analysis can be broadly divided into two strategies. The first
makes use of the hybrid method outlined in
\cite{Golterman:2014ksa}. The second uses continuous momenta in the
lattice Fourier transform to compute the scalar HVP function directly
at arbitrary momentum \cite{DelDebbio:2015a}.

\section{Simulation Details}

Simulations have been performed on the two 2+1 flavour domain wall
fermion (DWF) ensembles with near-physical pion masses described in
\cite{Blum:2014tka}. For convenience we summarize the properties of
these ensembles in Table \ref{tab:Ensembles}.
\begin{table}
  \protect\caption{\label{tab:Ensembles}Ensembles used in this study \cite{Blum:2014tka}.}

  \noindent \centering{}%
  \begin{tabular}{|c|c|c|}
    \hline 
    Parameter & 48I & 64I\\
    \hline 
    \hline 
    $L^{3}\times T\times L_{s}$ & $48^{3}\times96\times24$ & $64^{3}\times128\times12$\\
    \hline 
    $am_{l}$ & 0.00078 & 0.000678\\
    \hline 
    $am_{s}$ & 0.0362 & 0.02661\\
    \hline 
    $a^{-1}$ / GeV & 1.730(4) & 2.359(7)\\
    \hline 
    $L$ / fm & 5.476(12) & 5.354(16)\\
    \hline 
    $m_{\pi}$ / MeV & 139.2(4) & 139.2(5)\\
    \hline 
    $m_{K}$ / MeV & 499.0(12) & 507.6(16)\\
    \hline 
    $m_{\pi}L$ & 3.863(6) & 3.778(8)\\
    \hline
  \end{tabular}
\end{table}

We compute the lattice vacuum polarisation, $C_{\mu\nu}$, using
$\mathbb{Z}_2$ wall sources and M\"obius domain wall fermions, with a
local vector current at the source and the DWF conserved vector
current at the sink, i.e.:

\begin{equation}
  C_{\mu\nu}(x)=\frac{Z_V}{9}\left<{\cal V}_{\mu}(x)V_{\nu}(0)\right>,
\end{equation}

\noindent
where $Z_V$ is the vector renormalization constant, $a$ is the lattice
spacing, $V_\nu$ is the local vector current and we define the
conserved M\"obius DWF vector current ${\cal V}_{\mu}(x)$ as described
in \cite{Blum:2014tka}.

To account for a small mistuning in the strange quark mass on each
ensemble, we performed a set of partially quenched measurements using
the physical value of the strange quark mass. These were performed in
addition to the unitary measurements \cite{Blum:2014tka}.

\section{Analysis}

We implemented a variety of analysis strategies in order to ascertain
the dependence of $a_{\mu}$ on the analysis technique.

\subsection{HVP Computation}

We can compute the HVP tensor in momentum space by performing a
Fourier transform of the position space HVP correlator, i.e.:

\begin{equation}
  \Pi_{\mu\nu}\left(Q\right) = \sum_x {\rm e}^{-{\rm i} Q \cdot x}C_{\mu\nu}(x) - \sum_x C_{\mu\nu}\left(x\right),
\end{equation}

\noindent
where the second summation effectively subtracts the zero-mode
\cite{Bernecker:2011gh}. In the infinite volume limit this term is
zero, and subtracting it greatly reduces the noise in the low-$Q^2$
region. For the lowest momentum value of $\Pi\left(\hat{Q}^2\right)$
the improvement in the statistical error is approximately a factor of
five.

We then perform a tensor decomposition of the HVP tensor, so that it
may be related to the scalar HVP function as follows:

\begin{equation}
  \Pi_{\mu\nu}\left(\hat{Q}\right)=\left(\delta_{\mu\nu}\hat{Q}^{2}-\hat{Q}_{\mu}\hat{Q}_{\nu}\right)\Pi\left(\hat{Q}^{2}\right) + \cdots,
\end{equation}

\noindent
where the ellipis denotes contributions from Lorentz symmetry
breaking, discretisation and finite volume effects and
$\hat{Q} = 2 \sin\left(Q / 2\right)$ is the momentum of the
intermediate photon. We remove a potential source of lattice cut-off
effects by considering only the diagonal component of the HVP tensor
where $\hat{Q}_\mu = 0$ \cite{Boyle:2011hu}.

\subsection{Hybrid Method}

We used the hybrid method as described in \cite{Golterman:2014ksa}.
This method consists of partitioning the integrand in
(\ref{eq:amu-integral}) into three non-overlapping adjacent regions
using cuts at low- and high-$Q^{2}$. The integrand is then computed
for the three regions in different ways. The low-$Q^{2}$ region is
integrated by modelling $\Pi(Q^{2})$ to extrapolate to $\Pi(0)$, which
is subtracted to compute $\hat{\Pi}(Q^{2})$. This result is then
combined with the kernel $f(Q^{2})$ to produce the integrand of
interest, which is then integrated numerically. The mid-$Q^{2}$ region
is integrated directly by multiplying the lattice data by $f(Q^{2})$
before using the trapezium method. Finally, the high-$Q^{2}$ region is
integrated by using the result from perturbation theory
\cite{Chetyrkin:1996cf, Golterman:2014ksa}.  Restricting the use of an
HVP parameterisation to the low-$Q^2$ region allows us to minimise
systematic effects \cite{Golterman:2014ksa}.

We use two classes of parameterisations for the low-$Q^{2}$ region
when performing the integral in Equation (\ref{eq:amu-integral}): Padé
approximants and conformal polynomials. The Padé approximants are
written as follows \cite{Aubin:2012me}:

\begin{equation}
  R_{mn}\left(\hat{Q}^{2}\right)=\Pi_{0}+\hat{Q}^{2}\left(\sum_{i=0}^{m-1}\frac{a_{i}^{2}}{b_{i}^{2}+\hat{Q}^{2}}+\delta_{mn}c^{2}\right),\;n=m,\, m+1,
\end{equation}

\noindent
where $a_{i}$, $b_{i}$, $\Pi_{0}$ and possibly $c$ are parameters to
be determined.

The conformal polynomials are written as follows
\cite{Golterman:2014ksa}:

\begin{equation}
  P^E_{n}\left(\hat{Q}^{2}\right)=\Pi_{0}+\sum_{i=1}^{n}p_{i}w^{i},\; 
  w=\frac{1-\sqrt{1+z}}{1+\sqrt{1+z}},\; z=\frac{\hat{Q}^{2}}{E^{2}},
\end{equation}

\noindent
where $p_{i}$ and $\Pi_{0}$ are parameters to be determined. The
parameter $E$ is the two-particle mass threshold.

We use two techniques for constraining the low-$Q^{2}$ models:
$\chi^{2}$ minimisation and continuous time moments
\cite{Chakraborty:2014mwa}. The $\chi^{2}$ minimization involves a fit
where the covariance matrix is approximated by its diagonal, i.e. the
fit is uncorrelated. This technique lends weight to points in the
computed HVP with a smaller statistical error at larger values of
$Q^2$.

The moments method defines a relationship between the HVP scalar
function and the lattice space-averaged current-current correlator,
$C_{\mu\mu}(t)$.

\begin{equation}
  \sum_t {\rm e}^{-{\rm i}Q_0 t} C_{\mu\mu}(t) = \hat{Q}_0^2 \Pi\left(\hat{Q}_0^2\right)
\end{equation}

\noindent
Taking the $n$th derivate with respect to $\hat{Q}_0$ at $\hat{Q}_0=0$
allows us to write

\begin{equation}
  \left(-1\right)^{n}\sum_t t^{2n} C_{\mu\mu}(t) = \left. \frac{\partial^{2n}}{\partial Q_0^{2n}} \left( \hat{Q}_0^2 \Pi\left(\hat{Q}_0^2 \right)\right) \right|_{Q_0=0}
\end{equation}

\noindent
We then insert one of the above analytical ans\"atze for the HVP
scalar function, setting up a system of equations that can be solved
to determine the model parameters.

The moments method uses continuous derivatives, meaning an infinite
volume is assumed. When performing the moments method, we use a model
that is a function of $\hat{Q}^{2}$. However, within the moments
method, derivatives are taken with respect to $Q_0$ and not
$\hat{Q}_0$. Within the determination of the model parameters, the
low-$Q^2$ cut is not used as an input for this technique, so the
resulting parameters do not depend on the low cut used in the hybrid
method \cite{Chakraborty:2014mwa}.

\subsection{Continuous Momenta}

One alternative to the hybrid method is to compute the HVP directly at
an arbitrary momentum by performing the Fourier transform at said
momentum \cite{DelDebbio:2015a}. Whereas before we used
$Q_0=\frac{2\pi}{T}n_0$ with $n_{0}\in\mathbb{Z},\ -T/2\le n_{0}<T/2$,
we now let $n_{0}$ lie anywhere on the half-closed interval
$[-T/2,T/2)$. This allows for the computation of $a^{\rm s}_\mu$
without using a parameterisation of the HVP.

Because we are computing the HVP tensor for momenta that are
non-Fourier modes on the lattice, there may be some finite volume
errors associated with this method. However, it can be shown that
these are exponentially suppressed by the lattice volume
\cite{DelDebbio:2015a}.  Using this technique, we compute the HVP at
arbitrary momenta up to some high cut, after which the perturbative
result is used.

\section{Results}

We used nine different parameterisations of the HVP when performing
the hybrid method: $P_2^{0.5 {\rm GeV}}$, $P_3^{0.5 {\rm GeV}}$,
$P_4^{0.5 {\rm GeV}}$, $P_2^{0.6 {\rm GeV}}$, $P_3^{0.6 {\rm GeV}}$,
$P_4^{0.6 {\rm GeV}}$, $R_{0,1}$, $R_{1,1}$ and $R_{1,2}$. We scan
three low cuts and three high cuts: $0.5 {\rm GeV}^2$,
$0.7 {\rm GeV}^2$ and $0.9 {\rm GeV}^2$, and $4.5 {\rm GeV}^2$,
$5.0 {\rm GeV}^2$ and $5.5 {\rm GeV}^2$. We used the same high cuts
when computing $a_\mu^{\rm s}$ using continuous momenta, where we used
a step size of 0.005 for $n_t$.

\begin{figure}[ht]
  \centering
  \begin{tabular}{cc}
    \includegraphics[scale=0.5]{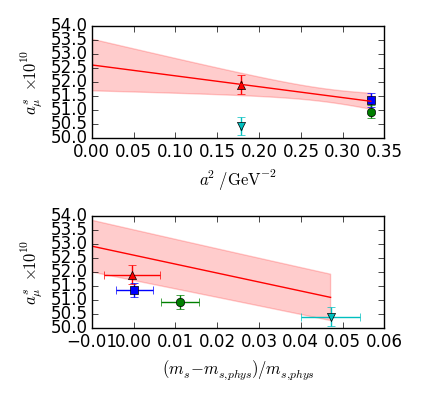} &
                                                                                               \includegraphics[scale=0.5]{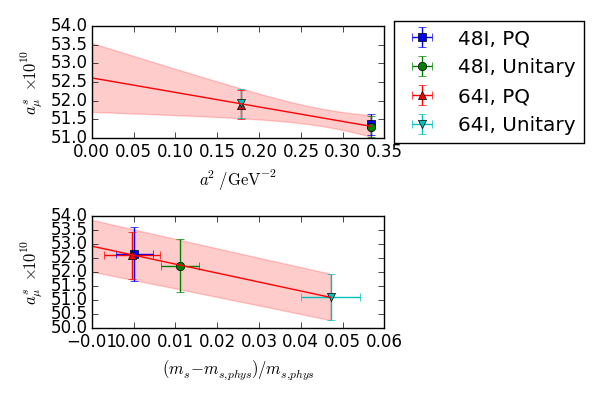} \\
                                                                                               (a) & (b) \\
  \end{tabular}
  \caption{\label{fig:extrapolations}Continuum and strange quark mass
    extrapolations. In the right-hand set of plots we have subtracted
    the effects from the strange quark mass (top) and lattice spacing
    (bottom).}
\end{figure}

Figure \ref{fig:extrapolations} illustrates an example extrapolation
to the continuum and the physical strange quark mass. We perform a
two-dimensional linear fit in $a^2$ and the relative deviation of the
strange mass from the physical value. We do this because domain wall
fermions are ${\cal O}(a)$ improved, and in the latter case we assume
a linear dependence of $a_\mu^{\rm s}$ on the strange quark mass. In
this case we used the $R_{0,1}$ parameterisation, which was
constrained using an uncorrelated $\chi^2$ minimisation. The low cut
in this case was $0.5 {\rm GeV}^2$ and the high cut was
$4.5 {\rm GeV}^2$. The effect of the strange quark mistuning is
clearly visible, with the final value of $a_\mu^{\rm s}$ shifting from
approximatley $50\times10^{-10}$ to $53.0\times10^{-10}$.

\begin{figure}[ht]
  \centering
  \begin{tabular}{cc}
    \includegraphics[scale=0.5]{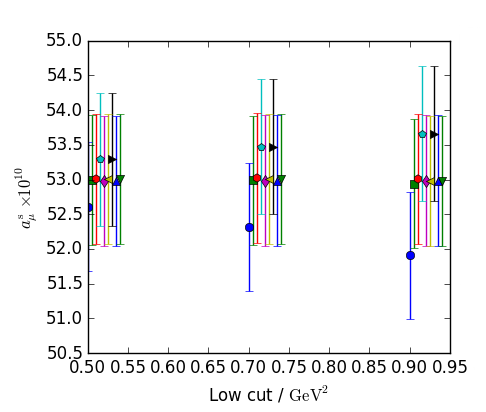} &
                                                                                               \includegraphics[scale=0.5]{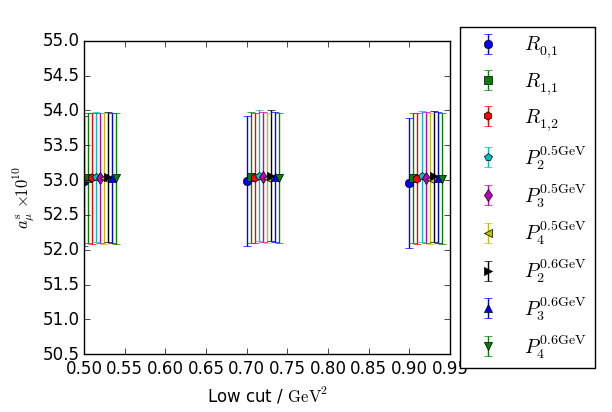} \\
                                                                                               (a) & (b) \\
  \end{tabular}
  \caption{\label{fig:values-of-amu-vs-cut}Computed values of
    $a_\mu^{\rm s}$ against various low cuts for fits (left) and
    moments (right).}
\end{figure}

Figure \ref{fig:values-of-amu-vs-cut} illustrates the variation of
$a^{\rm s}_\mu$ as the low cut in the hybrid method is varied. All the
computed values of $a_\mu^{\rm s}$ agree within statistics, and most
of the values are in strong agreement with one another. Furthermore,
our results agree with those of HPQCD \cite{Chakraborty:2014mwa} and
ETMC \cite{Feng:2013xsa} to within statistics. The models with the
fewest parameters, i.e. $P_2^{\rm 0.5 GeV}$, $P_2^{\rm 0.6 GeV}$ and
$R_{0,1}$, deviate slightly from $53.0\times 10^{-10}$.  This is more
apparent in the case where the models are constrained with $\chi^2$
fits. This is likely a result of the fit favouring data at larger
$Q^2$, where the statistical error is smaller, whilst the moments use
an expansion around $Q^2=0$, favouring data around this point.

\begin{figure}[ht]
  \centering
  \includegraphics[scale=0.7]{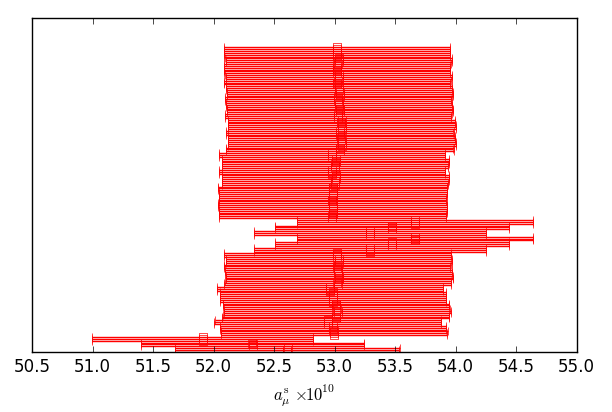}
  \caption{\label{fig:histogram}Errorbar plot illustrating the various
    values of $a_\mu$ computed in this analysis.}
\end{figure}

Figure \ref{fig:histogram} demonstrates the various values of $a_\mu$
computed in this analysis. Good agreement is found between all values
of $a_\mu^{\rm s}$. This suggests that the systematic error resulting
from the various analysis techniques is small.

\section{Summary}

We have computed the strange contribution to the anomalous magnetic
moment of the muon using domain wall fermions with physical quark
masses. We used a variety of analysis techniques, in particular the
hybrid method proposed in \cite{Golterman:2014ksa} and continuous
momenta \cite{DelDebbio:2015a}. Our final values of $a_\mu^{\rm s}$
show good agreement with each other, suggesting that the systematic
error from the choice of analysis technique is small.  Furthermore, we
find good agreement with the work of HPQCD \cite{Chakraborty:2014mwa}
and ETMC \cite{Feng:2013xsa}.

We are now in the process of finalising the analysis of possible
sources of systematic error, particularly finite volume effects. We
are simultaneously extending our analysis to the connected light
contribution contribution to $a_\mu$. In the future we plan to account
for the effect of disconnected diagrams.

\section{Acknowledgements}

This work is part of a programme of research by the RBC/UKQCD
collaboration. This research was funded by the European Research
Council under the European Community’s Seventh Framework Programme
(FP7/2007-2013) ERC grant agreement No 279757. The authors also
acknowledge STFC grants ST/J000396/1 and ST/L000296/1. M.S. is funded
by an EPSRC Doctoral Training Centre grant (EP/G03690X/1) through the
ICSS DTC. The calculations reported here have been done on the DiRAC
Bluegene/Q computer at the University of Edinburgh’s Advanced
Computing Facility.

\bibliographystyle{pos} \bibliography{sources}

\end{document}